\documentclass[sigconf,natbib=true,anonymous=false]{acmart}

\AtBeginDocument{%
  \providecommand\BibTeX{{%
    \normalfont B\kern-0.5em{\scshape i\kern-0.25em b}\kern-0.8em\TeX}}}
\usepackage{adjustbox}
\copyrightyear{2024}
\acmYear{2024}
\setcopyright{rightsretained}
\acmConference[SIGIR-AP '24]{Proceedings of the 2024 Annual International ACM SIGIR Conference on Research and Development in Information Retrieval in the Asia Pacific Region}{December 9--12, 2024}{Tokyo, Japan}
\acmBooktitle{Proceedings of the 2024 Annual International ACM SIGIR Conference on Research and Development in Information Retrieval in the Asia Pacific Region (SIGIR-AP '24), December 9--12, 2024, Tokyo, Japan}\acmDOI{10.1145/3673791.3698432}
\acmISBN{979-8-4007-0724-7/24/12}

\makeatletter
\gdef\@copyrightpermission{
   \begin{minipage}{0.3\columnwidth}
     \href{https://creativecommons.org/licenses/by-nc-sa/4.0/}{\includegraphics[width=0.90\textwidth]{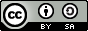}}
   \end{minipage}\hfill
   \begin{minipage}{0.7\columnwidth}
     \href{https://creativecommons.org/licenses/by-nc-sa/4.0/}{This work is licensed under a Creative Commons Attribution-ShareAlike International 4.0 License.}
   \end{minipage}
   \vspace{5pt}
}
\makeatother

\usepackage[inline]{enumitem}
\begin{document}

\title{A Reproducibility and Generalizability Study of Large Language Models for Query Generation}

\author{Moritz Staudinger}
\orcid{0000-0002-5164-2690}
\affiliation{%
  \institution{TU Wien}
  \city{Vienna}
  \country{Austria}
}
\email{moritz.staudinger@tuwien.ac.at}

\author{Wojciech Kusa}
\orcid{0000-0003-4420-4147}
\affiliation{%
  \institution{TU Wien \& Allegro ML Research}
  \city{Vienna, Austria \& Warsaw}
  \country{Poland}}
\email{wojciech.kusa@tuwien.ac.at}

\author{Florina Piroi}
\orcid{0000-0001-7584-6439}
\affiliation{%
  \institution{TU Wien}
  \city{Vienna}
  \country{Austria}}
\email{florina.piroi@tuwien.ac.at}

\author{Aldo Lipani}
\orcid{0000-0002-3643-6493}
\affiliation{%
  \institution{University College London}
  \city{London}
  \country{United Kingdom}}
\email{aldo.lipani@ucl.ac.uk}

\author{Allan Hanbury}
\orcid{0000-0002-7149-5843}
\affiliation{%
  \institution{TU Wien}
  \city{Vienna}
  \country{Austria}}
\email{allan.hanbury@tuwien.ac.at}

\renewcommand{\shortauthors}{Moritz Staudinger, Wojciech Kusa, Florina Piroi, Aldo Lipani, Allan Hanbury}

\newcommand{\wojciech}[1]{\textcolor{magenta}{Wojciech: #1}}

\newcommand{\todo}[1]{\textcolor{red}{todo: #1}}

\definecolor{customgray}{RGB}{90, 90, 90}
\newcommand{\customquote}[1]{\textcolor{customgray}{``\textit{#1}''}}

\begin{abstract}
Systematic literature reviews (SLRs) are a cornerstone of academic research, yet they are often labour-intensive and time-consuming due to the detailed literature curation process. 
The advent of generative AI and large language models (LLMs) promises to revolutionize this process by assisting researchers in several tedious tasks, one of them being the generation of effective Boolean queries that will select the publications to consider including in a review.
This paper presents an extensive study of Boolean query generation using LLMs for systematic reviews, reproducing and extending the work of \citet{wang2023chatgpt} and \citet{alaniz2023utility}. 
Our study investigates the replicability and reliability of results achieved using ChatGPT and compares its performance with open-source alternatives like Mistral and Zephyr to provide a more comprehensive analysis of LLMs for query generation.

Therefore, we implemented a pipeline, which automatically creates a Boolean query for a given review topic by using a previously defined LLM, retrieves all documents for this query from the PubMed database and then evaluates the results. 
With this pipeline we first assess whether the results obtained using ChatGPT for query generation are reproducible and consistent.
We then generalize our results by analyzing and evaluating open-source models and evaluating their efficacy in generating Boolean queries.

Finally, we conduct a failure analysis to identify and discuss the limitations and shortcomings of using LLMs for Boolean query generation.
This examination helps to understand the gaps and potential areas for improvement in the application of LLMs to information retrieval tasks.
Our findings highlight the strengths, limitations, and potential of LLMs in the domain of information retrieval and literature review automation. Our code is available online\footnote{\label{gitlink}\url{https://github.com/MoritzStaudinger/booleanQueryReproducibility}}.
\end{abstract}



\maketitle

\section{Introduction}

A systematic literature review (SLR) is a type of secondary study that summarizes available primary studies fitting pre-specified criteria to answer precise research questions. It uses rigorous scientific methods to minimize bias and generate clear, solid conclusions that health practitioners frequently use to make decisions \cite{Jo2009}.

Unfortunately, conducting systematic reviews is slow, labor intensive and time consuming, as this is primarily based on human effort. A recent estimate shows that conducting a complete systematic review takes, on average, 67 weeks \cite{Borah2017AnalysisRegistry}, although another previous study reports that the median time to publication was 2.4 years \cite{Tricco2008FollowingStudy}. Furthermore, according to \citet{Shojania2007HowAnalysis}, 23\% of the published systematic reviews need to be updated within two years after completion. 

Systematic reviews in medicine rely on constructing Boolean queries for conducting searches for relevant publications~\cite{hausner2015development,Higgins2019}.
The development and use of Boolean queries in systematic literature reviews addresses several critical challenges inherent to the review process. They ensure reproducibility, allowing other researchers to replicate searches and validate findings.
Search strategies are always reported when the systematic review is published.
Secondly, Boolean queries improve the efficiency and accuracy of literature searches by precisely defining inclusion and exclusion criteria.
This usually happens by maximizing the Recall with a sacrifice on the Precision, ensuring comprehensive search results.

Creating Boolean queries for SLRs is a critical yet complex process. It involves the formulation of search strategies using Boolean logic to combine keywords and phrases that define the inclusion and exclusion criteria for studies. This task requires not only a deep understanding of the topic at hand, but also proficiency in database search functionalities and syntax. The complexity arises from the need to balance specificity and sensitivity of the search: too broad a query may yield an unmanageable number of irrelevant results, while too narrow a query risks missing key studies. Moreover, constructing effective Boolean queries demands familiarity with synonyms, medical subject headings (MeSH\footnote{Medical Subject Headings: \url{https://www.ncbi.nlm.nih.gov/mesh}} terms in PubMed), and the various ways research questions can be articulated in the literature. This expertise is typically developed over years, making the task challenging for newcomers or those outside of library and information science disciplines. Consequently, the development of Boolean queries represents a significant bottleneck in the SLR process, contributing to its overall time-consuming nature \cite{Tsafnat2014SystematicTechnologies,OMara-Eves2015}.

With the initial release of ChatGPT by OpenAI, November 2022, a new era of general-purpose language generating assistants began\footnote{\url{https://openai.com/blog/chatgpt}}. 
Since then numerous companies, as Meta~\cite{touvron2023llama}, Mistral AI or Alphabet~\cite{team2023gemini}, announced and published their own large language models (LLM) to compete with OpenAI and allow researchers, companies and others to use it for speeding up their tasks.
While LLMs and specifically ChatGPT started to be used extensively by the research community, the reproduction of generated results remains a challenge. 
This is mainly due to many models being not disclosed to the public, and the only way of interacting with these tools is by using APIs for interacting with them. 
The ease of use of LLMs via API, as opposed to fine-tuning pre-trained language models (PLMs), has lowered the entry barrier for non-technical researchers. 
However, this accessibility could potentially pose a threat to research integrity and reproducibility~\cite{balloccu-etal-2024-leak}.

Although OpenAI and Mistral AI have extended their APIs to allow the configuration of a random seed, even this beta function does not guarantee a deterministic output and the reproduction of generated outputs\footnote{\url{https://community.openai.com/t/the-seed-option-for-gpt-does-not-increase-the-determinism-level/512892}}.
The inherent LLM output variability poses a challenge to the reproducibility of systematic reviews, necessitating rigorous validation of LLM-generated queries against expert strategies to ensure reliability and relevance~\cite{yun2023appraising}.
While models, such as Llama~\cite{touvron2023llama}, Alpaca~\cite{alpaca}, and Mistral~\cite{jiang2023mistral} are open-source, their performance in domain-specific tasks and low resource setting is heavily influenced by the original datasets used for their training. %

In this paper, we attempt to reproduce and generalize the results and extend the setup of two very recent publications on Boolean Query Generation with ChatGPT. \citet{wang2023chatgpt} published an influental work at SIGIR in 2023, which proposes using ChatGPT for Boolean Query Generation, and~\citet{alaniz2023utility} investigated Boolean Query Generation from a medical perspective using ChatGPT-4.

As of 5th July 2024, the paper by~\citet{wang2023chatgpt} has been cited 152 times according to Google Scholar and 111 times according to Semantic Scholar within less than one year since its publication, which is highly unusual in the domain of systematic literature review automation\footnote{For comparison, the first and one of the most highly cited papers in the domain of screening prioritisation by~\citet{Cohen2006} has been cited 350 times in 18 years since its publication.}.
Semantic Scholar classifies the intent of almost half of these citations as ``background'', meaning that this paper was used as a supportive evidence for (similar) projects.
With the existence of the publication by~\citet{alaniz2023utility}, it has become undeniable that medical experts also investigate the possibilities of using LLMs to improve systematic reviews.

Furthermore, \citet{guimaraes2024development} conducted a critical analysis of LLMs in developing search strategies for systematic reviews, underscoring the importance of transparent, reproducible, and standardized methods in clinical decision-making.
Despite ChatGPT's ability to simplify technical content, its shortcomings in accurately incorporating synonymous terms, organizing search terms, and excluding irrelevant information challenge its reliability. Furthermore, the absence of a filter for randomized clinical trials underscores the necessity for cautious application.
\citet{guimaraes2024development} emphasize the importance of not solely depending on ChatGPT or other LLM for search strategies, advocating for researcher and methodologist awareness of their limitations.
While their work discussed several important points from the perspective of practitioners, it did not analyze the computational and experimental issues with using LLMs for Boolean query generation.

Our work aims to cover these aspects by providing a more generalized view of this topic.
We study the following research questions:

\begin{itemize}
    \item[\textbf{RQ1}] To what extent are results of Boolean query generation reproducible and reliable, when repeatedly running the experiment? %
    \item[\textbf{RQ2}] How do open-source large language models perform in comparison to GPT models in the task of Boolean query creation?
    \item[\textbf{RQ3}] What are the shortcomings and limitations of using LLMs for Boolean query generation?
\end{itemize}

This paper is structured as follows: Section~\ref{sec:related} reviews related work, situating our study within the existing research landscape on Boolean query generation and LLMs in information retrieval tasks, while Section~\ref{sec:methodology} outlines our methodology, including baseline comparisons and SLR examples. Section~\ref{sec:results} details the results of our reproducibility and generalizability study.
In Section~\ref{sec:discussion} we discuss the implications of our results, and explore challenges encountered, including issues with Seed collection~\cite{Wang2022} and example queries. 
Lastly, Section~\ref{sec:findings} summarizes findings in response to our research questions.
Our implementation is available online\textsuperscript{\ref{gitlink}}.%

\section{Related Work} \label{sec:related}

This section presents existing studies on Boolean query creation, citation screening automation, and the application of instruction-based LLMs in IR and SLR automation.
\vspace{-0.5em}
\subsection{Boolean query formulation and refinement}
Query formulation involves deriving a Boolean query based on the research question, following specific guidelines. There are two main procedures for developing queries for systematic reviews. The first is the \emph{conceptual} method \cite{clark2013systematic}, which starts by identifying high-level concepts from pilot searches or known relevant studies. These concepts help discover synonyms and related keywords, leading to an iterative refinement process guided by the expertise of the information specialist.

The second procedure is the \emph{objective} method \cite{hausner2015development}, which begins by creating a small set of potentially relevant studies, similar to the conceptual method. Statistical procedures then extract terms from these studies, providing a weak validation of the query. These terms are manually added to the query, again guided by the expertise of the information specialist. Both of these methods are time-consuming, involving multiple pilot searches and validation steps \cite{karimi2009challenge}, and are prone to human error \cite{sampson2006errors}.

To address these challenges, \citet{Scells2021} explored automating these procedures. Their research found that while computational methods alone could not match human effectiveness, manual refinement of automatically generated queries significantly improved retrieval effectiveness. Automatic query refinement emerged from the benefits of manual refinement. These methods take an initial query and apply transformations (adding terms or clauses) to enhance effectiveness \cite{10.1093/jamia/ocaa148,scells2018searchrefiner}. Combined with query visualization tools \cite{scells2018searchrefiner}, these refinement tools can improve the initial query.

\vspace{-0.5em}
\subsection{Document screening automation} %

Alongside Boolean query improvements, citation screening automation has emerged as a crucial component in improving the systematic review process. 
Research has explored various dimensions, including the use of diverse data sources \cite{Lee2018,Scells2020,Kusa2022AutomationStudy,kusa2023cruise}, the employment of different ranking algorithms and models \cite{Alharbi2017,Chen2017,Kanoulas2018,Minas2018,Wang2022,Wang2023b}, and the implementation of active learning strategies to enhance screening efficiency via a human-in-the-loop approach \cite{Anagnostou2017,Carvallo2020,DiNunzio2017,Hollmann2017,Norman2018,Singh2017,Yang2022,kusa2023outcome}. 

\vspace{-0.5em}
\subsection{Instruction-based large language models} %

Recent advancements in instruction-based LLMs, exemplified by ChatGPT, demonstrate these models' capability to accurately interpret and execute user instructions across diverse tasks \cite{Gozalo-Brizuela2023,Guo2023,Sallam2023,jiang2024mixtral,jiang2023mistral}. With their massive parameter count and extensive training on varied textual data, these models can generate relevant and coherent responses to a wide array of inquiries \cite{Gozalo-Brizuela2023}. Studies have assessed the efficacy of ChatGPT and other instruction-based LLMs, such as Alpaca, in tasks like query generation for systematic reviews, showcasing their potential to reach or outperform previous methodologies \cite{Omar2023,Tan2023,wang2023chatgpt}. 

Furthermore, the integration of instruction-based LLMs with ranking models has led to more effective outcomes in information retrieval tasks, employing strategies like retrieval-then-generation and generation-then-retrieval to leverage the strengths of both model types for enhanced query understanding and document retrieval \cite{Gao2022,Jiang2023,Wang2023,Yasunaga2023}.

\section{Methodology} \label{sec:methodology}

We describe, now, the methodology for our reproducibility study. 

\vspace{-0.5em}
\subsection{Datasets}

Following \citet{wang2023chatgpt} we use the Seed~\cite{Wang2022} and CLEF TAR~\cite{Kanoulas2017,Kanoulas2018} collections, and use the same set of topics.
We obtain the datasets from the original GitHub repositories\footnote{\url{https://github.com/ielab/sysrev-seed-collection}}$^{,}$\footnote{\url{https://github.com/CLEF-TAR/tar}}.
Due to significant overlaps with the CLEF TAR collection, we refrain from using the CSMeD~\cite{kusa2023csmed} and SIGIR17~\cite{scells2017test} datasets.
Unfortunately, there are no other datasets that provide search queries for systematic review automation~\cite{kusa2023csmed}.

\vspace{-0.5em}
\subsection{Baselines}

Three types of baseline queries for the CLEF TAR and the Seed datasets are mentioned in the study by \citet{wang2023chatgpt}, called \textit{original}, \textit{conceptual}, and \textit{objective}.
Beyond stating them in the paper, \citet{wang2023chatgpt} do not explain what kind of queries they are and how they were constructed.
We assume that the \textit{original} baseline represents the original Boolean query created by the human experts when conducting the systematic review.
However, ~\citet{wang2023chatgpt} do not describe the specific approach to constructing the \textit{conceptual} and \textit{objective} baselines.
The three related papers they refer to in this respect,~\cite{clark2013systematic,simon2010identifying,hausner2015development}, are limited to describing the general theory on how the conceptual and objective methods are used by information specialists to create the Boolean queries.
We found no information about these approaches in the original dataset papers~\cite{Kanoulas2017,Kanoulas2018}.

For the CLEF TAR dataset, we were not able to reproduce the baselines, as this dataset does not provide queries in the PubMed API compatible format, and the paper did not state how the original queries were created. 
Therefore, we opted to copy the baseline results provided in the original study.

\vspace{-0.5em}
\subsection{Models}

\citet{wang2023chatgpt} state that they used ChatGPT for their experiments. Based on the submission deadline of their paper (January 31st 2023), and the release of GPT-4 (March 13th 2023), we assume that it was GPT 3.5 that was used for the experiments. 
\citet{alaniz2023utility} stated that they used ChatGPT 4 for their experiments.

To generalize their results, we followed their zero-shot generation approach.
We select the following GPT models from the OpenAI API: GPT-3.5-1106, GPT-3.5-0125 and GPT-4-1106 (\textit{GPT-4}).
We also select two open-source models using the Mistral API Mistral-tiny (Mistral-7B-v0.2) and Mistral-small (Mixtral-8X7B-v0.1)\footnote{\url{https://docs.mistral.ai/api/}}:.

To obtain more reliable results, we ran each generation prompt with ten different seeds for the smaller models (GPT-3.5-1106, Mistral~\cite{jiang2023mistral}, Mixtral~\cite{jiang2024mixtral}), and five different seeds for GPT-3.5-0125 and GPT-4.
As both publications did not mention how they issued the prompts to ChatGPT, we opted to use the OpenAI and Mistral AI APIs, and compare the results obtained in our experiments with those obtained by~\citet{wang2023chatgpt} and \citet{alaniz2023utility}.

Additionally to the available APIs, we run two open source LLMs locally, namely
Mistral-7B-Instruct-v0.2\footnote{\url{https://huggingface.co/mistralai/Mistral-7B-Instruct-v0.2}} (\textit{Mistral-local}) and Zephyr-7b-beta\footnote{\url{https://huggingface.co/HuggingFaceH4/zephyr-7b-beta}} (\textit{Zephyr}), to validate the results of the closed-source APIs with publicly available models. We picked these two models, as both are instruction-tuned-based models, which are trained to fulfill tasks based on given input data. 

\vspace{-0.5em}
\subsection{Approaches to query generation}

In their work, \citet{wang2023chatgpt} generated Boolean queries by issuing three different levels of zero-shot prompts (denoted as queries \textbf{q1}, \textbf{q2}, \textbf{q3}), two different one-shot prompts (\textbf{q4} and \textbf{q5}), two different query refinement prompts (\textbf{q6} and \textbf{q7}) and one guided prompt generation approach to ChatGPT (\textbf{guided} query). This guided approach consists of four steps, similar to a Chain-of-Thought prompting approach: \begin{enumerate*}[label*=(\arabic*)]
\item the relevant terms are created based on SLR Title and Abstract;
\item the terms are, then, classified into \textit{medical}, \textit{treatment} and \textit{design} terms;
\item the terms and their classes are then composed to an initial Boolean query;
\item in the end the terms are enriched with MeSH fields to improve the performance.
\end{enumerate*}

For their experiments ~\citet{alaniz2023utility} also used a guided prompt design, which differs from ~\citet{wang2023chatgpt} by combining (1) and (2), while expanding (4) to increase the usage of MeSH terms.

\citet{wang2023chatgpt} ran each query once, and measured the variability of their queries, by rerunning their best prompt ten times. 
\citet{alaniz2023utility} performed a case study based on a single systematic review, with five validation examples for intermediary steps. While they showed the generalizability of their generated queries for their subdomain, they did not analyze the reliability of their approach or compare their approach with any baseline.

\begin{figure}
    \centering
    \includegraphics[width=\columnwidth]{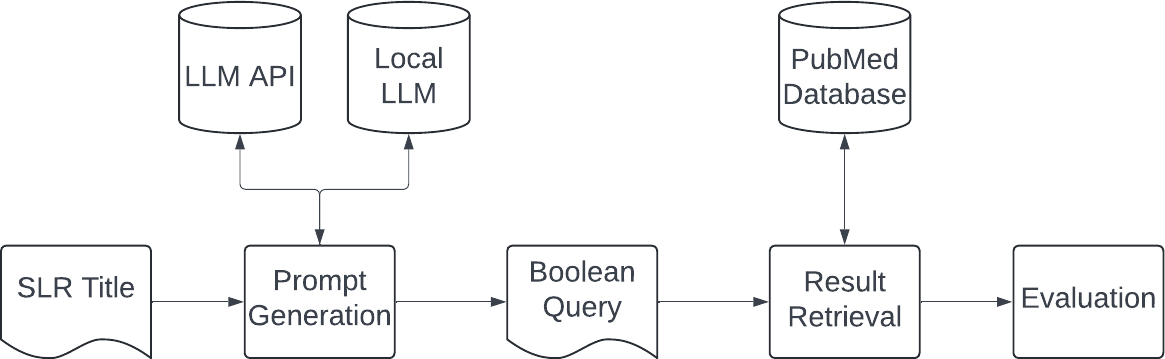}
    \caption{Proposed pipeline to generate and evaluate boolean queries}
    \label{fig:pipeline}
    \vspace{-0.5cm}
\end{figure}

For this work, we implemented the pipeline shown in Figure~\ref{fig:pipeline} to generate queries based on an input SLR title and a defined LLM, and then execute the query on the PubMed database over their API and evaluate the results on the defined metrics.

We used \citet{wang2023chatgpt} given prompts for query generation, but refined the input prompts based on the LLM used and the desired output. We instructed the Mistral models not to include any filtering based on years, and to only return the Boolean query.
For the GPT models, we used the available \textit{system} role, and split the original created prompts into system and user prompts, to assist the model in understanding the task. 
\vspace{-0.5em}
\subsection{Query examples} \label{sec:query-examples}
Looking at queries \textbf{q4} and \textbf{q5}, \citet{wang2023chatgpt} measure the effectiveness of two types of examples:  (1) one high-quality systematic review example (\textbf{HQE}), and (2) an example that is similar or related to the querying topic (\textbf{RE}).
We use the same HQE as in the original paper (review CD010438 from CLEF TAR collection).

For finding RE, we follow a similar path, which we describe below.
We use the PubMedBERT model\footnote{\url{https://huggingface.co/pritamdeka/S-PubMedBert-MS-MARCO}} in the dense encoder setting using SentenceTransformers~\cite{reimers-2019-sentence-bert} and \texttt{retriv} Python packages instead of the MonoBERT approach. 
We then use a similarity score, to extract the relevant PubMed queries from the Seed collection. 

Regarding the CLEF TAR dataset, the strategy to obtain the correct PubMed queries, which are not part of the CLEF TAR dataset, was not described. We tried to reconstruct the authors' high-quality review with the ID CD010438, by using Polyglot\footnote{\url{https://sr-accelerator.com/\#/polyglot}} to translate the CLEF TAR query to their used example query, but we did not succeed in creating a similar valid query. 
Thus we could not run the relevant example queries q4 and q5 for the CLEF TAR dataset.

\vspace{-0.5em}
\subsection{Machine-actionable LLM output}

\citet{wang2023chatgpt} state: \customquote{After obtaining Boolean queries from ChatGPT, any incorrectly formatted Boolean queries were removed and generated again for the review topic} (p 4)\footnote{\customquote{Styled text} denotes a quote from one of the original papers: \citep{wang2023chatgpt} or \citep{alaniz2023utility}.}.

While \citet{wang2023chatgpt} mention that they re-generated incorrectly formatted queries, \citet{alaniz2023utility} did not elaborate on their approach for such situations of formally incorrect queries.
As ChatGPT responses are usually not in a standardized format and don't contain additional information (see examples in Table \ref{tab:example_outputs} and the discussion in Section~\ref{sec:query_generation}), and neither of the two publications' authors specified how the valid Boolean queries were extracted, we can only speculate on how this was done. 
In our work, to allow the automatic extraction of the Boolean query, we adapted the prompt and the API call, so that the GPT model returns a predefined JSON object. 
For the remaining models we execute the generated output ``as is'', i.e. without any post-processing.

\vspace{-0.5em}
\subsection{Evaluation}

For their evaluation, \citet{wang2023chatgpt} computed \customquote{set-based measures such as precision, f-measure, and recall to assess the retrieved PubMed IDs using the abstract-level relevant assessment in the collections.} (p. 4).
Specifically they evaluate Precision, Recall, F1-score, and F3-score.
The F3-score is a variation of the F-measure that gives higher importance to Recall, which is the more important aspect of the SLR task, rather than Precision. 
In order to achieve comparable results, we follow their approach.
In the paper, we present Precision, Recall, and F1-score results. 
Due to space constraints, results for the F3-score are reported in the code repository.

\section{Results} \label{sec:results}

We first discuss the results of our reproducibility and generalizability study.
Then we present our results of the query and model variability and detailed analysis of LLM-generated Boolean queries.

\begin{table*}[t]
    \caption{Obtained average Precision scores with standard deviation for each model and for each prompt type (query). Statistically significantly better performing results (one sample t-test, $p < 0.05$) compared to \citet{wang2023chatgpt} are indicated by $\dag$. \textbf{Bolded} results are the highest overall, and \underline{underlined} results are the highest for a given query.}
    \label{tab:precisionResults}
\begin{adjustbox}{max width=0.94\textwidth}
\begin{tabular}{@{}lrrrrrrrr@{}}
\toprule
\textbf{CLEF}                    & \multicolumn{8}{c}{\textbf{Precision}}                                                                                                                                                                                                                                                                                                    \\ \midrule
Baseline                         & \multicolumn{8}{c}{0.021}                                                                                                                                                                                                                                                                                                               \\ \midrule
                                 & \multicolumn{1}{c}{\textbf{\citet{wang2023chatgpt}}} & \multicolumn{1}{c}{\textbf{GPT-3.5-1106}} & \multicolumn{1}{c}{\textbf{GPT-3.5-0125}} & \multicolumn{1}{c}{\textbf{GPT-4}} & \multicolumn{1}{c}{\textbf{Mistral-tiny}} & \multicolumn{1}{c}{\textbf{Mistral-small}} & \multicolumn{1}{c}{\textbf{Mistral-local}} & \multicolumn{1}{c}{\textbf{Zephyr}} \\ \midrule
\multicolumn{1}{l|}{q1}          & 0.054                                & 0.345 ± 0.257$^{\dag}$                        & 0.064 ± 0.173                             & \underline{0.362 ± 0.262}$^{\dag}$                      & 0.227 ± 0.274                             & 0.303 ± 0.263                              & 0.209 ± 0.291                              & 0.062 ± 0.164                       \\
\multicolumn{1}{l|}{q2}          & 0.117                                & 0.319 ± 0.274$^{\dag}$                        & 0.072 ± 0.181                             & \underline{0.350 ± 0.248}$^{\dag}$                       & 0.233 ± 0.297                             & 0.144 ± 0.242                              & 0.281 ± 0.304                              & 0.122 ± 0.230                        \\
\multicolumn{1}{l|}{q3}          & 0.084                                & 0.320 ± 0.257$^{\dag}$                         & 0.070 ± 0.181                              & \underline{\textbf{0.378 ± 0.244}}$^{\dag}$                      & 0.249 ± 0.293                             & 0.258 ± 0.275                              & 0.148 ± 0.220                               & 0.071 ± 0.153                       \\
\multicolumn{1}{l|}{q4-HQE}          & 0.075                                & 0.284 ± 0.264$^{\dag}$                        & 0.044 ± 0.138                             & \underline{0.330 ± 0.270}$^{\dag}$                        & 0.114 ± 0.216                             & 0.160 ± 0.238                               & 0.099 ± 0.214                              & 0.184 ± 0.248                       \\
\multicolumn{1}{l|}{q5-HQE}          & 0.096                                & 0.290 ± 0.255$^{\dag}$                         & 0.050 ± 0.151                              & \underline{0.355 ± 0.269}$^{\dag}$                      & 0.104 ± 0.205                             & 0.181 ± 0.252                              & 0.129 ± 0.211                              & 0.055 ± 0.171                       \\ \midrule \midrule
\textbf{Seed}                    & \multicolumn{8}{c}{\textbf{Precision}}                                                                                                                                                                                                                                                             \\ \midrule
Baseline                         & \multicolumn{8}{c}{0.034}                                                                                                                                                                                                                                                                       \\
Baseline-edit                   & \multicolumn{8}{c}{0.035}                                                                                                                                                                                                                                                                       \\ \midrule
                                 & \multicolumn{1}{c}{\textbf{\citet{wang2023chatgpt}}} & \multicolumn{1}{c}{\textbf{GPT-3.5-1106}} & \multicolumn{1}{c}{\textbf{GPT-3.5-0125}} & \multicolumn{1}{c}{\textbf{GPT-4}} & \multicolumn{1}{c}{\textbf{Mistral-tiny}} & \multicolumn{1}{c}{\textbf{Mistral-small}} & \multicolumn{1}{c}{\textbf{Mistral-local}} & \multicolumn{1}{c}{\textbf{Zephyr}} \\ \midrule
\multicolumn{1}{l|}{q1}          & \underline{0.050}                                & 0.037 ± 0.094                        & 0.017 ± 0.037                             & 0.035 ± 0.139                      & 0.040 ± 0.120                               & 0.018 ± 0.053                              & 0.019                                      & 0.012                               \\
\multicolumn{1}{l|}{q2}          & \underline{0.098}                                & 0.006 ± 0.071                        & 0.010 ± 0.032                              & 0.029 ± 0.100                        & 0.030 ± 0.123                              & 0.021 ± 0.125                              & 0.035                                      & 0.003                               \\
\multicolumn{1}{l|}{q3}          & \underline{0.073}                                 & 0.036 ± 0.160                         & 0.013 ± 0.039                             & 0.037 ± 0.135                      & 0.031 ± 0.121                             & 0.027 ± 0.091                              & 0.007                                      & 0.000                                 \\
\multicolumn{1}{l|}{q4-HQE}          & \underline{0.028}                                & 0.027 ± 0.107                        & 0.010 ± 0.030                               & 0.012 ± 0.048                      & 0.019 ± 0.090                              & 0.007 ± 0.038                              & 0.003                                      & 0.010                                \\
\multicolumn{1}{l|}{q5-HQE}          & 0.019                                & 0.037 ± 0.140$^{\dag}$                         & 0.011 ± 0.024                             & 0.010 ± 0.035                       & 0.024 ± 0.104                             & 0.007 ± 0.030                               & \underline{0.044}                                      & 0.009                               \\ \midrule
\multicolumn{1}{l|}{q4-RE} & \underline{0.035}                                & 0.023 ± 0.061                        & 0.009 ± 0.027                             & 0.013 ± 0.074                      & 0.021 ± 0.104                             & 0.008 ± 0.041                              & 0.028                                      & 0.021                               \\
\multicolumn{1}{l|}{q5-RE} &     ---                                  & 0.024 ± 0.095                        & 0.018 ± 0.084                             & 0.015 ± 0.044                      & \underline{0.025 ± 0.105}                             & 0.011 ± 0.043                              & \underline{0.025}                                      & 0.010                                \\
\multicolumn{1}{l|}{guided}                    & \textbf{\underline{0.099}}                                & 0.006 ± 0.071                        &         0.009 ± 0.073                                  & 0.006 ± 0.029                      & 0.006 ± 0.059                             & 0.012 ± 0.065                              &              ---                              &    ---                                 \\ \bottomrule
\end{tabular}
\end{adjustbox}
\end{table*}

\begin{table*}[t]
    \caption{Obtained average F1 scores with standard deviation for each model and for each prompt type (query). Statistically significantly better performing results (one sample t-test, $p < 0.05$) compared to \citet{wang2023chatgpt} are indicated by $\dag$. \textbf{Bolded} results are the highest overall, and \underline{underlined} results are the highest for a given query. }
    \label{tab:f1Results}
\begin{adjustbox}{max width=0.94\textwidth}
\begin{tabular}{@{}lrrrrrrrr@{}}
\toprule
\textbf{CLEF}                      & \multicolumn{8}{c}{\textbf{F1-score}}                                                                                                                                                                                                                                                                   \\ \midrule
Baseline                           & \multicolumn{8}{c}{0.029}                                                                                                                                                                                                                                                                       \\ \midrule
                                   & \multicolumn{1}{l}{\textbf{\citet{wang2023chatgpt}}} & \multicolumn{1}{l}{\textbf{GPT-3.5-1106}} & \multicolumn{1}{l}{\textbf{GPT-3.5-0125}} & \multicolumn{1}{l}{\textbf{GPT-4}} & \multicolumn{1}{l}{\textbf{Mistral-tiny}} & \multicolumn{1}{l}{\textbf{Mistral-small}} & \multicolumn{1}{l}{\textbf{Mistral-local}} & \multicolumn{1}{l}{\textbf{Zephyr}} \\ \midrule
\multicolumn{1}{l|}{q1}            & 0.050                                  & \underline{0.097 ± 0.139}$^{\dag}$                        & 0.021 ± 0.078                             & 0.085 ± 0.13$^{\dag}$                       & 0.048 ± 0.104                             & 0.064 ± 0.107                              & 0.040                                     & 0.014                              \\
\multicolumn{1}{l|}{q2}            & 0.065                                & 0.084 ± 0.124$^{\dag}$                        & 0.021 ± 0.079                             & \underline{0.096 ± 0.14}$^{\dag}$                       & 0.034 ± 0.084                             & 0.028 ± 0.074                              & 0.033                                     & 0.021                              \\
\multicolumn{1}{l|}{q3}            & 0.044                                & \underline{0.101 ± 0.131}$^{\dag}$                        & 0.025 ± 0.088                             & 0.097 ± 0.129$^{\dag}$                      & 0.046 ± 0.107                             & 0.070 ± 0.119                               & 0.031                                     & 0.010                              \\
\multicolumn{1}{l|}{q4-HQE}            & 0.064                                & \underline{0.123 ± 0.161}$^{\dag}$                        & 0.027 ± 0.100                               & 0.086 ± 0.132$^{\dag}$                      & 0.050 ± 0.120                               & 0.064 ± 0.119                              & 0.022                                     & 0.051                              \\
\multicolumn{1}{l|}{q5-HQE}            & 0.072                                & \underline{\textbf{0.134 ± 0.160}}$^{\dag}$                         & 0.024 ± 0.087                             & 0.102 ± 0.133$^{\dag}$                      & 0.038 ± 0.098                             & 0.057 ± 0.104                              & 0.045                                     & 0.007                              \\ \midrule \midrule
\textbf{Seed}                      & \multicolumn{8}{c}{\textbf{F1-score}}                                                                                                                                                                                                                                                                   \\ \midrule
Baseline                           & \multicolumn{8}{c}{\textbf{0.060}}                                                                                                                                                                                                                                                                       \\
Baseline-edit                     & \multicolumn{8}{c}{0.058}                                                                                                                                                                                                                                                                       \\ \midrule
                                   & \multicolumn{1}{l}{\textbf{\citet{wang2023chatgpt}}} & \multicolumn{1}{l}{\textbf{GPT-3.5-1106}} & \multicolumn{1}{l}{\textbf{GPT-3.5-0125}} & \multicolumn{1}{l}{\textbf{GPT-4}} & \multicolumn{1}{l}{\textbf{Mistral-tiny}} & \multicolumn{1}{l}{\textbf{Mistral-small}} & \multicolumn{1}{l}{\textbf{Mistral-local}} & \multicolumn{1}{l}{\textbf{Zephyr}} \\ \midrule
\multicolumn{1}{l|}{q1}            & 0.027                                & \underline{0.038 ± 0.081}$^{\dag}$                        & 0.026 ± 0.051                             & 0.023 ± 0.060                       & 0.033 ± 0.078                             & 0.027 ± 0.071                              & 0.024                                     & 0.015                               \\
\multicolumn{1}{l|}{q2}            & \underline{0.031}                                 & 0.003 ± 0.018                        & 0.015 ± 0.042                             & 0.025 ± 0.066                      & 0.021 ± 0.069                             & 0.010 ± 0.042                               & 0.026                                     & 0.007                              \\
\multicolumn{1}{l|}{q3}            & \underline{0.033}                                & 0.019 ± 0.053                        & 0.018 ± 0.044                             & 0.024 ± 0.062                      & 0.022 ± 0.073                             & 0.032 ± 0.087                              & 0.012                                      & 0.000                                   \\
\multicolumn{1}{l|}{q4-HQE}            & \underline{0.027}                                & 0.023 ± 0.053                        & 0.016 ± 0.043                             & 0.017 ± 0.055                      & 0.014 ± 0.041                             & 0.008 ± 0.030                               & 0.007                                     & 0.021                              \\
\multicolumn{1}{l|}{q5-HQE}            & 0.019                                & \underline{0.032 ± 0.074}                        & 0.019 ± 0.037                             & 0.015 ± 0.048                      & 0.016 ± 0.045                             & 0.009 ± 0.029                              & 0.020                                       & 0.015                              \\ \midrule
\multicolumn{1}{l|}{q4-RE}   & 0.014                                 & \underline{0.030 ± 0.067}$^{\dag}$                         & 0.015 ± 0.041                             & 0.012 ± 0.047                      & 0.013 ± 0.041                             & 0.010 ± 0.047                               & \underline{0.030}                                     & 0.019                              \\
\multicolumn{1}{l|}{q5-RE}   &  ---                                     & \underline{0.026 ± 0.077}                        & 0.018 ± 0.044                             & 0.021 ± 0.062                      & 0.017 ± 0.045                             & 0.013 ± 0.046                              & 0.024                                     & 0.015                              \\
\multicolumn{1}{l|}{guided} & \underline{0.049}                                & 0.002 ± 0.011                        &         0.004 ± 0.019                                  & 0.009 ± 0.035                      & 0.004 ± 0.028                             & 0.012 ± 0.042                              &                 ---                           &        ---                             \\ \bottomrule
\end{tabular}
\end{adjustbox}
\end{table*}
\vspace{-0.5em}

\begin{table*}[t]
    \caption{Obtained average Recall scores with standard deviation for each model and for each prompt type (query). Statistically significantly better performing results (one sample t-test, $p < 0.05$) compared to \citet{wang2023chatgpt} are indicated by $\dag$. \textbf{Bolded} results are the highest overall, and \underline{underlined} results are the highest for a given query.}
    \label{tab:recallResults}
\begin{adjustbox}{max width=0.94\textwidth}
\begin{tabular}{@{}lrrrrrrrr@{}}
\toprule
\textbf{CLEF}                      & \multicolumn{8}{c}{\textbf{Recall}}                                                                                                                                                                                                                                                               \\ \midrule
Baseline                           &  \multicolumn{8}{c}{\textbf{0.832}}                                                                                                                                                                                                                                                                        \\ \midrule
                                   & \multicolumn{1}{c}{\textbf{\citet{wang2023chatgpt}}} & \multicolumn{1}{c}{\textbf{GPT-3.5-1106}} & \multicolumn{1}{c}{\textbf{GPT-3.5-0125}} & \multicolumn{1}{c}{\textbf{GPT-4}} & \multicolumn{1}{c}{\textbf{Mistral-tiny}} & \multicolumn{1}{c}{\textbf{Mistral-small}} & \multicolumn{1}{c}{\textbf{Mistral-local}} & \multicolumn{1}{c}{\textbf{Zephyr}} \\ \midrule
\multicolumn{1}{l|}{q1}            & \underline{0.129}                                & 0.084 ± 0.145                        & \multicolumn{1}{r}{0.019 ± 0.079}         & \multicolumn{1}{r}{0.072 ± 0.142}  & \multicolumn{1}{r}{0.046 ± 0.119}         & \multicolumn{1}{r}{0.062 ± 0.128}          & \multicolumn{1}{r}{0.037}                 & \multicolumn{1}{r}{0.015}          \\
\multicolumn{1}{l|}{q2}            &\underline{0.131}                                 & 0.067 ± 0.125                        & \multicolumn{1}{l}{0.019 ± 0.081}         & \multicolumn{1}{r}{0.093 ± 0.169}  & \multicolumn{1}{r}{0.026 ± 0.071}         & \multicolumn{1}{r}{0.021 ± 0.063}          & \multicolumn{1}{r}{0.025}                 & \multicolumn{1}{r}{0.017}           \\
\multicolumn{1}{l|}{q3}            & \underline{0.118}                                & 0.115 ± 0.195                        & \multicolumn{1}{r}{0.026 ± 0.106}         & \multicolumn{1}{r}{0.086 ± 0.147}  & \multicolumn{1}{r}{0.041 ± 0.108}         & \multicolumn{1}{r}{0.063 ± 0.123}          & \multicolumn{1}{r}{0.038}                  & \multicolumn{1}{r}{0.007}          \\
\multicolumn{1}{l|}{q4-HQE}            & \underline{0.504}                                & 0.139 ± 0.212                        & \multicolumn{1}{r}{0.033 ± 0.131}         & \multicolumn{1}{r}{0.086 ± 0.170}   & \multicolumn{1}{r}{0.063 ± 0.166}         & \multicolumn{1}{r}{0.067 ± 0.150}           & \multicolumn{1}{r}{0.020}                 & \multicolumn{1}{r}{0.062}          \\
\multicolumn{1}{l|}{q5-HQE}            & \underline{0.334}                                & 0.150 ± 0.212                         & \multicolumn{1}{r}{0.027 ± 0.107}         & \multicolumn{1}{r}{0.091 ± 0.142}  & \multicolumn{1}{r}{0.043 ± 0.129}         & \multicolumn{1}{r}{0.060 ± 0.138}           & \multicolumn{1}{r}{0.053}                 & \multicolumn{1}{r}{0.005}          \\ \midrule \midrule
\textbf{Seed}                      & \multicolumn{8}{c}{\textbf{Recall}}                                                                                                                                                                                                                                                               \\ \midrule
Baseline                           &  \multicolumn{8}{c}{\textbf{0.711}}                                                                                                                                                                                                                                                                       \\
Baseline-edit                     &   \multicolumn{8}{c}{0.647}                                                                                                                                                                                                                                                                       \\ \midrule
                                   & \multicolumn{1}{c}{\textbf{\citet{wang2023chatgpt}}} & \multicolumn{1}{c}{\textbf{GPT-3.5-1106}} & \multicolumn{1}{c}{\textbf{GPT-3.5-0125}} & \multicolumn{1}{c}{\textbf{GPT-4}} & \multicolumn{1}{c}{\textbf{Mistral-tiny}} & \multicolumn{1}{c}{\textbf{Mistral-small}} & \multicolumn{1}{c}{\textbf{Mistral-local}} & \multicolumn{1}{c}{\textbf{Zephyr}} \\ \midrule
\multicolumn{1}{l|}{q1}            & \multicolumn{1}{r}{0.053}            & \multicolumn{1}{r}{0.148 ± 0.24$^{\dag}$}     & \underline{0.203 ± 0.291}$^{\dag}$                             & 0.132 ± 0.25$^{\dag}$                       & 0.132 ± 0.244                             & 0.190 ± 0.293                               & 0.122                                     & 0.036                              \\
\multicolumn{1}{l|}{q2}            & \multicolumn{1}{r}{0.039}            & \multicolumn{1}{r}{0.025 ± 0.108}    & \underline{0.147 ± 0.247}$^{\dag}$                             & 0.141 ± 0.246$^{\dag}$                      & 0.047 ± 0.142                             & 0.057 ± 0.152                              & 0.063                                     & 0.087                              \\
\multicolumn{1}{l|}{q3}            & \multicolumn{1}{r}{0.052}            & \multicolumn{1}{r}{0.086 ± 0.206$^{\dag}$}    & 0.169 ± 0.255$^{\dag}$                             & 0.156 ± 0.244$^{\dag}$                      & 0.065 ± 0.168                             & \underline{0.214 ± 0.299}                              & 0.095                                     & 0.002                              \\
\multicolumn{1}{l|}{q4-HQE}            & \multicolumn{1}{r}{0.129}             & \multicolumn{1}{r}{0.213 ± 0.310$^{\dag}$}     & \underline{0.237 ± 0.296}$^{\dag}$                             & 0.145 ± 0.267                      & 0.091 ± 0.184                             & 0.092 ± 0.228                              & 0.121                                     & 0.193                              \\
\multicolumn{1}{l|}{q5-HQE}            & \multicolumn{1}{r}{0.079}            & \multicolumn{1}{r}{0.244 ± 0.311$^{\dag}$}    & \underline{0.258 ± 0.326}$^{\dag}$                             & 0.113 ± 0.224$^{\dag}$                      & 0.064 ± 0.155                             & 0.084 ± 0.210                               & 0.092                                     & 0.153                              \\ \midrule
\multicolumn{1}{l|}{q4-RE}   & \multicolumn{1}{r}{0.016}            & \multicolumn{1}{r}{0.174 ± 0.256$^{\dag}$}    & \underline{0.202 ± 0.280}$^{\dag}$                              & 0.088 ± 0.205$^{\dag}$                      & 0.080 ± 0.171                              & 0.080 ± 0.212                               & 0.055                                     & 0.066                              \\
\multicolumn{1}{l|}{q5-RE}   & \multicolumn{1}{r}{---}                  & \multicolumn{1}{r}{0.178 ± 0.281}    & \underline{0.267 ± 0.340}                              & 0.104 ± 0.205                      & 0.067 ± 0.160                              & 0.075 ± 0.195                              & 0.064                                     & 0.060                              \\
\multicolumn{1}{l|}{guided} & \multicolumn{1}{r}{\underline{0.517}}            & \multicolumn{1}{r}{0.035 ± 0.130}     &                   0.048 ± 0.109                        & 0.125 ± 0.221                      & 0.017 ± 0.090                              & 0.100 ± 0.206                                &     ---                                       &    ---                                 \\ \bottomrule
\end{tabular}
\end{adjustbox}
\end{table*}
\subsection{Reproducibility and Generalizability Study}
As described above, we run our queries repeatedly with fixed random seeds in the attempt to reproduce the results of \citet{wang2023chatgpt}. 
For both the CLEF TAR and the Seed dataset, we used the provided Boolean query as a baseline.
However, since the Seed dataset also contains a second Boolean query \textit{edited-search}, we also show the results for this query (\textit{Baseline-edit} in the result tables), as it is unclear which query has been used based on the documentation.

Looking at the Precision scores (Table \ref{tab:precisionResults}) for all different queries generated for the CLEF TAR Dataset, we see that the GPT-3.5-1106 with a max Precision of $0.345$ and the GPT-4 model with a max Precision of $0.378$ both significantly outperform the results we were trying to reproduce ($0.096$). The Mistral API models (Mistral-tiny and Mistral-small) and also the locally run Mistral model outperform the original results on this dataset as well.
In contrast, for six out of eight query types on the Seed dataset, we couldn't reproduce the original results, and none of our models reached the scores reported by~\citet{wang2023chatgpt}.

We confirm the findings of \citet{wang2023chatgpt}, who stated that their GPT model outperforms the state of the art baselines with regard to F-measure on the CLEF TAR dataset, but not on the Seed dataset. 
Our results for F1-score (Table~\ref{tab:f1Results}) also indicate that GPT-3.5-1106, GPT-4, and Mistral-small, all outperform the baseline on multiple generated queries. 
A similar conclusion can be drawn based on the F3-score results.

When looking at the Recall, as displayed in Table \ref{tab:recallResults}, we can see that the all our models outperform the presented results on the Seed dataset by the authors, but perform worse on the CLEF TAR dataset. 
Nevertheless, based on our results we agree with the opinion of~\citet{wang2023chatgpt} that LLM generated queries are not suitable for high-recall retrieval, as even our best performing queries are significantly worse than the baseline. 

In general we can say that our tested LLMs generally perform better for precision scores on the CLEF TAR dataset and better for Recall values on the Seed dataset, when comparing our results with the authors' results. 
However, we do not see any clear trend whether the generation is more Recall or Precision oriented.
 
The guided approach of~\citet{wang2023chatgpt} is especially interesting as none of our tested LLMs managed to achieve a similar performance in any metric, when comparing it with their guided approach. They stated that their Precision and Recall results varied greatly and ranged from 0 to 1 for some of the topics.
\subsection{Variability}

\begin{figure}
    \centering
    \includegraphics[width=\columnwidth]{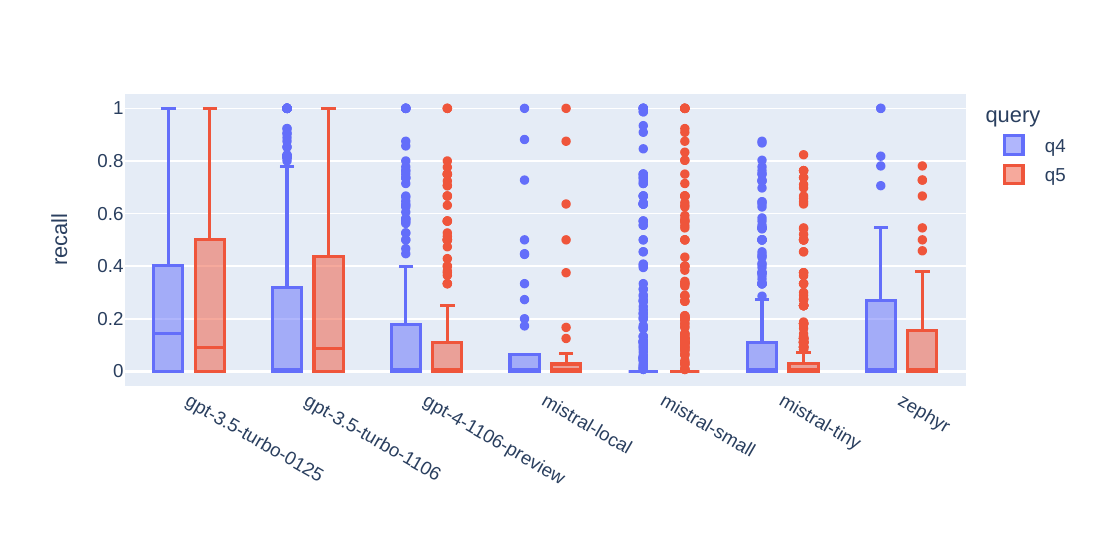}
    \caption{Recall variability for queries q4, q5 for the Seed dataset.}
    \label{fig:recallPlot}
\end{figure}

\begin{figure}
    \centering
    \includegraphics[width=\columnwidth]{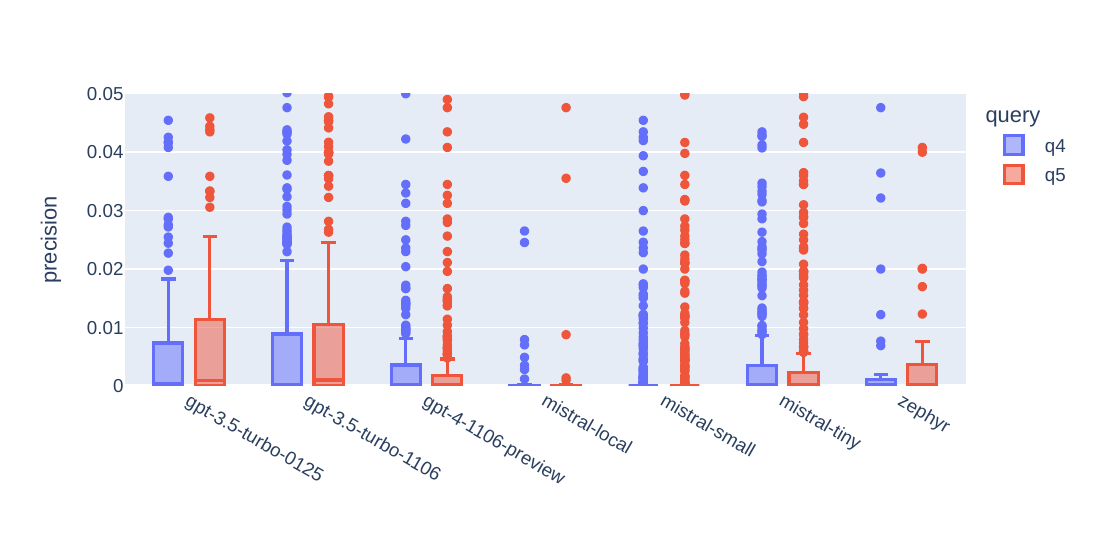}
    \caption{Precision variability for queries q4, q5 for the Seed dataset.}
    \label{fig:precisionPlot}
\vspace{-0.5em}
\end{figure}

\citet{wang2023chatgpt} state: \customquote{We show the variability of the query refinement in Figure 2. There is less variance in query refinement than in query formation (Figure 1)}.
We measure the variability in Recall and Precision for two of the best-performing queries: q4 and q5.
We compare the average Recall and Precision scores across topics and random seeds for the seven considered LLMs.
Figure~\ref{fig:recallPlot} presents the Recall results, while Figure~\ref{fig:precisionPlot} presents the results for Precision.
Both GPT-3.5 models consistently outperform other models in both evaluation measures however they also exhibit higher variability in generated Boolean queries.
Interestingly, Mistral-local exhibits lower variance and performs worse than Mistral-tiny despite both of them being based on the same Mistral-7B-v0.2 architecture.
\subsection{Query analysis}
\citet{wang2023chatgpt} mentioned that they regenerated incorrectly formatted queries, but did not specify how often this was done. 
Therefore, we conducted an in-depth analysis of the queries generated by the LLMs in our pipeline. 
\begin{figure}
    \centering
    \includegraphics[width=\columnwidth]{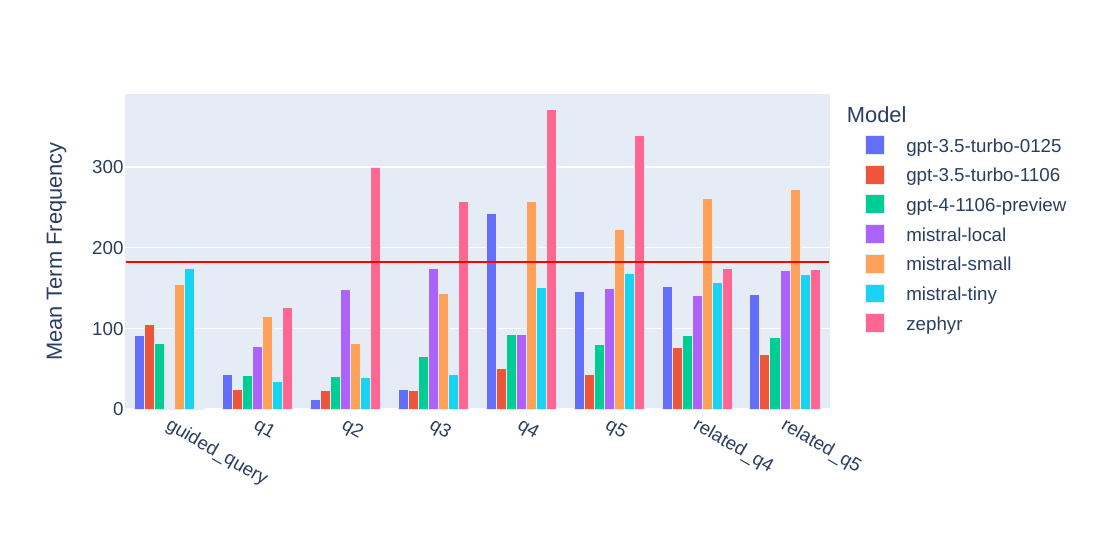}
    \caption{Average Boolean query term length per query for the \textit{Seed} dataset depending on the model. Red line denotes average for the expert-crafted Boolean queries.}
    \label{fig:booleanQueryTerms}
\end{figure}
\begin{figure}
    \centering
    \includegraphics[width=\columnwidth]{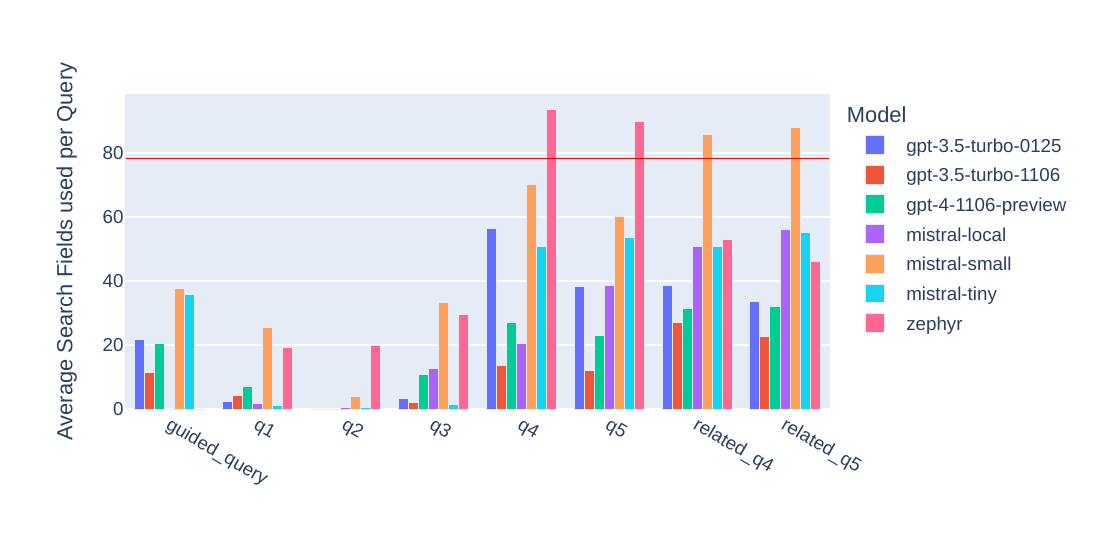}
    \caption{Used PubMed search fields per Boolean query. Red line denotes average for the expert-crafted Boolean queries.}
    \label{fig:correctSearchFields}
\end{figure}
Figure \ref{fig:booleanQueryTerms} shows the average number of terms the generated Boolean queries contain. All models, except Mistral-small and Zephyr, generate far fewer terms on average than the expert-crafted Boolean queries (red line). While Mistral-small outputs comparable, meaningful Boolean query terms for the q4 and q5 query types, Zephyr's output Boolean query terms are repeated multiple times (e.g. term `animal testing' appears over 160 times).
Nevertheless, the usage of PubMed search fields increases for all models, when they are provided with valid examples of search queries, using these fields (see Figure \ref{fig:correctSearchFields}).

\begin{figure}
    \centering
    \includegraphics[width=\columnwidth]{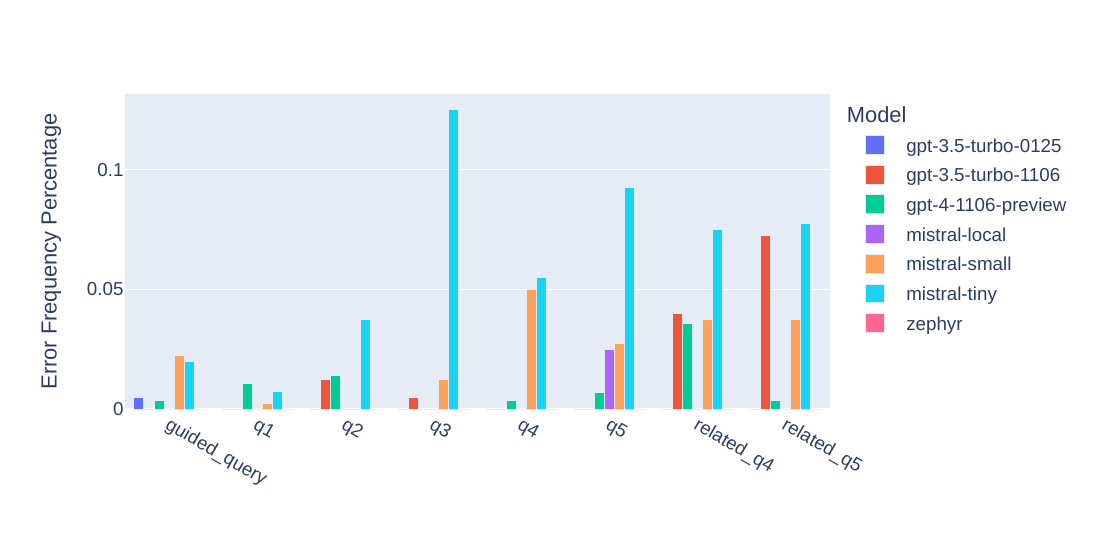}
    \caption{Percentage of incorrect query generations, per query type, for the \textit{Seed} dataset.}
    \label{fig:generationErrors}
\end{figure}

In Figure \ref{fig:generationErrors}, we see the percentage of errors that occured when generating the queries. In total $346$ out of $11,200$ generated queries were not created due to ReadTimeouts in Mistral AI or incorrect JSON generation in GPT.
Furthermore, in the \textbf{guided} approach, GPT-3.5-1106 was not always able to follow the communication and did not generate 181 out of around 400 queries (45\%).

\begin{figure}
    \centering
    \includegraphics[width=\columnwidth]{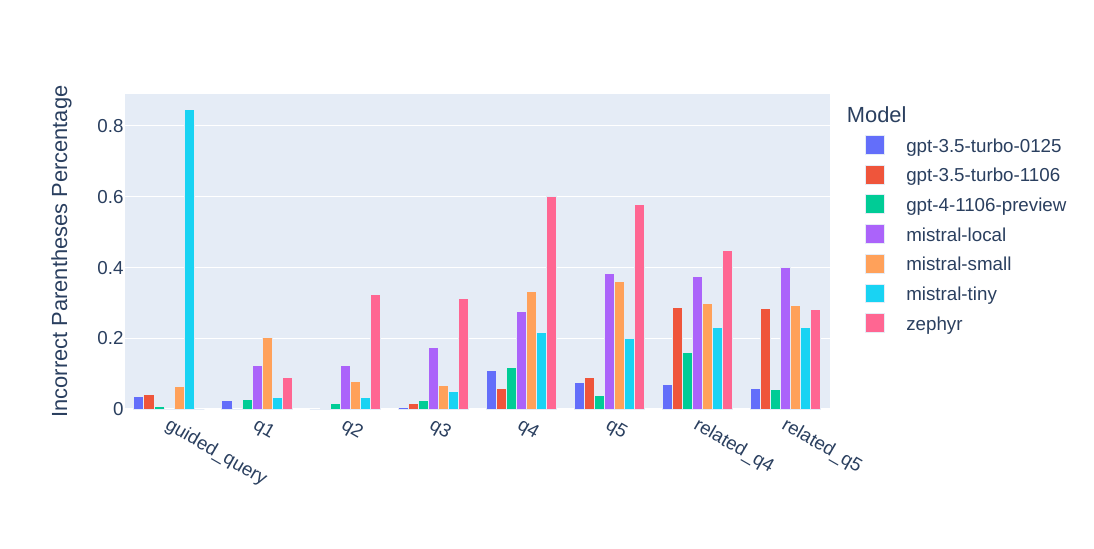}
    \caption{Percentage of wrongly generated parentheses, per query type, for the \textit{Seed} dataset.}
    \label{fig:incorrectParentheses}
\end{figure}
Further, the models did not manage to always set the parentheses correctly. In $1,771$ of $11,200$ generated queries, the parentheses were set incorrectly. Figure \ref{fig:incorrectParentheses} shows the percentage of wrongly generated parentheses. Similar to the generation errors from GPT-3.5-1106, mistral-tiny did not manage to follow the guided approach throughout and generated 328 wrong queries.

\section{Discussion} \label{sec:discussion}
While extending the setups by~\citet{wang2023chatgpt} and \citet{alaniz2023utility}, we ran into several issues and were unable to fully reproduce the publications, as not enough information was given by the authors. In this section, we want to discuss these issues.

\vspace{-0.5em}
\subsection{Query generation} \label{sec:query_generation}

The description of the query generation process used in the original papers is limited. 
While both groups of authors state the issued prompts with all necessary parameters, they neither describe how the prompts have been issued nor provide the source code to rerun their experiment. 
It is further not described how they extracted only the Boolean queries from the ChatGPT responses. 

At the time when the work by \citet{wang2023chatgpt} was published, there were no options for manipulating the output types from the ChatGPT responses nor for requesting only a specific output. At the time when \citet{alaniz2023utility} was published, these options were already available but are not discussed. 
From our experience, even when explicitly asked not to generate anything else except for the Boolean query, GPT models often return unnecessary explanations or other boilerplate texts (see Table~\ref{tab:example_outputs}).
We found that the only consistent way of generating the Boolean query only was to use the \texttt{JSON} return mode.
Parsing the output, either by extracting it through the API or from the website, caused issues because the generated outputs varied. These variations included additional linebreaks, comments before and after the query, and no dedicated character to separate the query and the text. \citet{alaniz2023utility} show in their Case Study all of the outputs for the different steps, which indicates that the queries were manually extracted.

{
\footnotesize
\begin{table*}[h]
    \centering
    \caption{Difference in raw outputs of the GPT-3.5 model for the same input prompt, random seed, \textit{top\_k=0.1} and \textit{temperature=0} depending on the model version and the return mode. Example SLR ID 43 from the \textit{Seed} collection. The original manually created Boolean query is in the first row. `\#Docs' column represents the number of publications returned after date filtering. The sentence in the gray color inside the example prompt was issued to models without \texttt{JSON} return mode.}
    \label{tab:example_outputs}
    \begin{tabular}{cc|p{0.6\textwidth}|ccc}
    \toprule
        Model & Return mode & \multicolumn{1}{|c|}{Raw model output} & \#Docs & Precision & Recall\\ \midrule
        \multicolumn{2}{c|}{Original Boolean query} & (("Thyroid Neoplasms"[MeSH] OR "Adenocarcinoma, Follicular"[MeSH] OR "Adenocarcinoma, Papillary"[MeSH] OR OPTC OR ((Thyroid[tiab] OR Follicular[tiab] OR Papillary[tiab] OR hurtle cell[tiab]) AND (cancer[tiab] OR cancers[tiab] OR carcinoma[tiab] OR carcinomas[tiab] OR Adenocarcinoma[tiab] OR Adenocarcinomas[tiab] OR neoplasm[tiab] OR neoplasms[tiab] OR nodule[tiab] OR nodules[tiab] OR tumor[tiab] OR tumour[tiab] OR Tumors[tiab] OR Tumours[tiab] OR cyst[tiab] OR cysts[tiab]))) AND ("Autopsy"[MeSH] OR "Autopsy"[tiab] OR "Autopsies"[tiab] OR "Postmortem"[tiab] OR Post-mortem[tiab] OR “step-sectioned”[tiab] OR “step sectioned”[tiab] OR  (Post[tiab] AND mortem[tiab])) AND ("Prevalence"[MeSH] OR "Prevalence"[tiab] OR "Prevalences"[tiab] OR Incidence[tiab] OR Epidemiology[tiab] OR Epidemiological[tiab] OR Frequency[tiab] OR Detected[tiab]) AND ("Incidental Findings"[MeSH] OR Incidental[tiab] OR Unsuspected[tiab] OR Discovery[tiab] OR Discoveries[tiab] OR Findings[tiab] OR Finding[tiab] OR Occult[tiab] OR Hidden[tiab] OR Latent[tiab] OR Consecutive[tiab])) & 198 & 0.131 & \textbf{0.765} \\ \midrule \midrule
        \multicolumn{2}{c|}{Models input prompt} & For a systematic review titled "\texttt{\{\{title\}\}}", can you generate a systematic review Boolean query to find all included studies on PubMed for the review topic? \textcolor{gray}{Structure the output as a JSON with the field boolean\_query and create the boolean query without filtering based on the year.} & \\ \midrule
        GPT3.5-turbo & --- & Sure! Here is a systematic review Boolean query to find all included studies on PubMed for the review topic "Prevalence of Differentiated Thyroid Cancer in Autopsy Studies Over Six Decades: A Meta-Analysis":

("Differentiated Thyroid Cancer" OR "Thyroid Carcinoma" OR "Papillary Thyroid Cancer" OR "Follicular Thyroid Cancer" OR "Hürthle Cell Thyroid Cancer" OR "Thyroid Neoplasms") AND ("Autopsy" OR "Postmortem" OR "Necropsy") AND ("Prevalence" OR "Incidence" OR "Epidemiology") AND ("Meta-Analysis" OR "Systematic Review") & 3 & 0.000 & 0.000
\\ \midrule
        GPT3.5-1106 & --- & Sure, here's a systematic review Boolean query for your topic:

("differentiated thyroid cancer" OR "papillary thyroid cancer" OR "follicular thyroid cancer" OR "thyroid carcinoma") AND ("autopsy" OR "postmortem" OR "cadaver" OR "necropsy") AND ("prevalence" OR "incidence" OR "epidemiology") 

You can use this query in PubMed to find relevant studies for your systematic review. & 78 & \textbf{0.154} & 0.353 
\\ \midrule
        GPT3.5-1106 & \texttt{JSON} & \{
  "boolean\_query": "(("thyroid cancer"[MeSH Terms] OR "thyroid neoplasms"[MeSH Terms] OR "thyroid carcinoma"[MeSH Terms] OR "differentiated thyroid cancer"[Title/Abstract]) AND (autopsy[Title/Abstract] OR autopsy studies[Title/Abstract]))" 
\}  & 3 & 0.000 & 0.000 \\ \midrule
        GPT3.5-0125 & --- & Sure! Here is a Boolean query that you can use to search for all included studies on PubMed for the systematic review titled "Prevalence of Differentiated Thyroid Cancer in Autopsy Studies Over Six Decades: A Meta-Analysis":

(("Differentiated Thyroid Cancer" OR "Thyroid Neoplasms") AND ("Autopsy" OR "Postmortem") AND ("Prevalence" OR "Incidence") AND ("Meta-Analysis" OR "Systematic Review")) & 370 & 0.057 & 0.618 \\ \midrule
        GPT3.5-0125 & \texttt{JSON}  & \{
  "boolean\_query": "(("differentiated thyroid cancer"[MeSH Terms] OR "thyroid neoplasms"[MeSH Terms] OR "thyroid cancer" OR "thyroid carcinoma") AND (autopsy[MeSH Terms] OR autopsy) AND ("prevalence" OR "incidence" OR "epidemiology") AND (meta-analysis[Publication Type] OR meta-analysis))"
\} & 3 & 0.000 & 0.000 \\ \bottomrule
    \end{tabular}
\end{table*}
}

\vspace{-0.5em}
\subsection{Example information}
While \citet{wang2023chatgpt} described large parts of their experiment in great detail, some of the key information to reproduce some of their experiments is missing. 
For the query refinement tasks q6 and q7, the authors state that they are using the \textit{original}, \textit{conceptual} and \textit{objective} refinement methods to improve the performance of the Boolean queries, but the meaning of this terms is not explained. 
Furthermore, the authors do not provide any information on how they obtained the Boolean queries.
This is particularly important when working with the CLEF TAR dataset, which does not contain queries in a format accepted by the PubMed API. 
Therefore, it would be necessary to manually translate the provided queries with a tool such as Polyglot, or obtain the original queries via an alternative way.
We were further confused by what the authors meant by \customquote{For query refinement, we identify three succesfull[sic] topics and eleven topics for query refinement}.
\vspace{-0.5em}
\subsection{Dataset issues}
We have encountered different kinds of reproducibility issues both for Seed and CLEF TAR collections, as well as the data used by \citet{alaniz2023utility} which we summarise below.

\subsubsection{Seed collection}

\customquote{The basis of our collection comes from our co-author at the Bond Institute for Evidence-Based Healthcare, who is a senior information
specialist. Each topic is a systematic review created using Boolean
queries developed by our co-author over the past five years. We
were provided unstructured data that we organised into 40 topics.}

We conducted an analysis of the \textit{collection.jsonl} file.
We found that the Seed dataset contains duplicated systematic reviews (the same publication URL with different queries. 
For example, topics 43 and 96 have the same title: ``Prevalence of Differentiated Thyroid Cancer in Autopsy Studies Over Six Decades: A Meta-Analysis''\footnote{\url{https://pubmed.ncbi.nlm.nih.gov/27601555/}} and share the same relevant documents.
Other examples with same title but different search name are (42, 51, 52 and 53 -- different included studies) and (7, 67 -- same included studies).

Moreover, we found that two SLRs have the same query, edited query, seed studies and search dates: (8 and 112) with slightly different SLR titles, meaning that these reviews are most probably the same.
Review (8) has one duplicated included study.
Despite the same input search parameters, these two topics have only 4 relevant documents in common, while 20 relevant documents are disjunct for each topic.

In total, it seems like the collection contains only 34 unique SLRs and not 40 as stated by the authors.
This might be problematic in the context of guided queries with examples where the authors used the title of the review and the query for prompting the LLM (q4-RE, Section~\ref{sec:query-examples}). 
From our findings, these duplicates would be the relevant examples for each other in all cases and after such prompting LLMs generate the query equal to the example review.

Another issue we found, is that the Seed dataset contains two different Boolean queries for each systematic review. When reproducing their baselines, we used both of these fields, \textit{edited-search} and \textit{query}, and compared them with the provided metrics of the original paper. 
Both of them differ from their provided baselines, but are reasonably close to the stated results.

\subsubsection{CLEF TAR}
\citet{wang2023chatgpt} describe the usage of this collection as: \customquote{We use the CLEF TAR 2017 and 2018 collections as they both consist of only diagnostic test accuracy (DTA) systematic reviews. Some topics from CLEF TAR 2017 are duplicated in the 2018 collection,so we obtained 72 review topics after removing these duplicates.}

We conducted the analysis of the CLEF TAR dataset\footnote{\url{https://github.com/CLEF-TAR/tar}}, and found that there are 80 review topics available in total from the 2017 and 2018 datasets. Eight of these topics\footnote{CD010771, CD007431, CD010772, CD010775, CD010783, CD010860, CD010896, CD011145} are part of the 2017 collection, but have been deleted in the 2018 collection, as they were unreliable~\cite{Kanoulas2018CLEFOverview}.
We conclude, thus, that \citet{wang2023chatgpt} used the 2018 collection, and not the 2017 collection.

It is further important to mention that both the CLEF TAR and the Seed datasets are publicly available on Github, and therefore closed-source LLMs could be aware of the given Systematic Reviews.

\subsubsection{Data by Alaniz}
While \citet{wang2023chatgpt} used already well-established datasets with known baselines, \citet{alaniz2023utility} used only a non-existing systematic review to display their results. They described their generalizability and verification of the steps as follows: \customquote{To confirm the generalizability across diverse topics, we implemented our process on existing systematic
reviews published in Plastic and Reconstructive Surgery}.

This evaluation is limited, as it only qualitatively compares the author created output with the output by the LLM, and does not make use of any pre-existing baselines or Boolean queries.

\section{Summary of Findings} \label{sec:findings}

In this Section, we summarize the findings for our RQs.

\begin{itemize}
    \item[\textbf{RQ1:}] \textit{To what extent are results of Boolean query generation reproducible and reliable, when rerunning the experiment?}
\end{itemize}

When comparing our results, with the results presented by \citet{wang2023chatgpt}, we saw similar trends to them, but were unable to reproduce their results. In general, we found that our results are neither reproducible nor reliable when rerunning the same prompts with different seeds, which led to a high variance in the obtained results. With the results fluctuating considerably, we do not consider any of the obtained results reliable enough for the critical task of systematic reviews.

\begin{itemize}
    \item[\textbf{RQ2:}] \textit{How do open-source large language models perform in comparison to GPT models in the task of Boolean query creation?}
\end{itemize}

We found that open-source LLMs perform reasonably well compared to commercial GPT models.
When looking at longer context lengths, we found that Open-Source models, due to them being smaller, are more prone to errors. This is especially eminent when looking at Mistral 7B for the guided query approach, with over 80 percent wrongly generated parentheses.

\begin{itemize}
    \item[\textbf{RQ3:}] \textit{What are the shortcomings and limitations of using LLMs for Boolean query generation?}
\end{itemize}

We extensively examined the shortcomings of automatic Boolean query generation via LLMs. We found that models do include more search field information in their output if they know about their existence (see, for example, the last row of Table~\ref{tab:example_outputs}). Therefore q1-q3 contain fewer search fields, than q4 and q5 which provide an example with possible search fields.
Additionally, not all models are capable of managing large conversation histories effectively, this issue stems from shorter context length for these models.

To obtain all the presented results, we spent 120 USD in API calls for GPT 3.5 and GPT 4, and 30 USD in API calls for Mistral AI. Thus, the total LLM-cost of our reproducibility study was 150 USD. Crafting candidate queries and, therefore, reducing the work necessary for experts is an upside, as this would reduce the overall cost of systematic reviews.

\section{Conclusion}
In conclusion, this research offers a critical examination of the role that generative AI and LLMs can play in streamlining the systematic literature review process through effective Boolean query generation. 
Despite the promising capabilities of these technologies, our findings reveal mixed results in terms of replicability and comparative performance of commercial models against open-source counterparts like Mistral and Zephyr. 
We also found and enumerated multiple reproducibility challenges related to the experimental setup concerning datasets, baselines and input used for prompts.
Challenges in reproducing the original studies by \citet{wang2023chatgpt} and \citet{alaniz2023utility} underscore the dynamic nature of LLMs and the importance of detailed documentation for replication studies. Furthermore, our study points out the limitations and the need for further exploration into the nuances of LLM-generated Boolean queries. 
As we navigate the evolving landscape of AI in academic research, this study serves as a step towards understanding the benefits and boundaries of integrating LLMs into the systematic review process.
However, challenges persist in ensuring the explainability of LLM-generated queries, necessitating the development of interpretability frameworks to shed light on their decision-making processes.
Based on our analysis we believe that using LLMs without proper fine-tuning do not give generalizable results for the Boolean query generation task. 
Therefore, we believe that future work has to make use of specially trained LLM to improve the quality of the generated queries and reduce the error rate on generated queries. Further work can also involve the usage of retrieval-augmented generation techniques to validate the correctness and the scope of generated queries.

\bibliographystyle{ACM-Reference-Format}
\bibliography{references}

\end{document}